\documentclass[fleqn,twoside]{article}%
\usepackage{amsmath}
\usepackage{amsfonts}
\usepackage{amssymb}
\usepackage{graphicx}
\usepackage{bm}% bold math

\newcommand{\be}{\begin{equation}}
\newcommand{\ee}{\end{equation}}
\newcommand{\bey}{\begin{eqnarray}}
\newcommand{\eey}{\end{eqnarray}}
\def\bes{\begin{equation}\begin{split}&}
\newcommand{\bi}{\bibitem}

\title{The issue of Branched Hamiltonian in F(T) Teleparallel Gravity}
\author{Manas Chakrabortty \footnote{E-mail:manas.chakrabortty001@gmail.com}, Kaushik Sarkar \footnote{E-mail: sarkarkaushik.rng@gmail.com}~ and Abhik Kumar Sanyal\footnote{E-mail: sanyal\_ ak@yahoo.com}.\\
~~~~~~~~~~~~~~~\\
$^*, ^\dag$ Dept. of Physics, Bankura University, Bankura, India - 722155,\\
$^\ddag$ Dept. of Physics, Jangipur College, Murshidabad, India - 742213\\}
\begin{document}
\maketitle
\begin{abstract}

\noindent
As in the case of Lanczos-Lovelock gravity, the main advantage of $F(T)$ gravity is said to be that it leads to second order field equations, while $F(R)$ gravity theory leads to fourth order equations. We show that it is rather a disadvantage, since it leads to the unresolved issue of `Branched Hamiltonian'. The problem is bypassed in $F(R,T)$ gravity theory.\\
\end{abstract}
\maketitle

\section{Introduction}
In the recent years, in analogy to the $F(R)$ theory of gravity, yet another extended theory of gravity, dubbed as `Teleparallel gravity' has drawn lot of attention. It is a generalized version of the so-called `Teleparallel gravity' originally proposed by Einstein \cite{E}. Einstein's attempt was to unify gravity and electromagnetism, going beyond the Riemannian metric. He characterized the concept of `direction’, `equality of directions’ or the so-called `parallelism’ for finite distances introducing a vierbein field along with the concept of absolute parallelism or Teleparallelism. In Teleparallel gravity, the curvature-less Weitzenb\"ock connection \cite{W} is considered, rather than the torsion-less Levi-Civita connection, which is used in General Relativity. Although, $F(T)$ Teleparallel theory of gravity was revived to drive inflation \cite{1}, later, it was proposed to drive the current accelerated expansion of our universe without considering dark energy \cite{2a,2b,2c,2d}. A comprehensive review of $F(T)$ teleparallel theory of gravity is available in the literature \cite{3}. \\

To consider Teleparallelism, the orthonormal tetrad components $e_{C} (x^{\alpha})$ \cite{4,5}, where the index $C$ runs over $0, 1, 2, 3$, are employed to the tangent space at each point $x^{\alpha}$ of the manifold. Their relation to the metric $g_{\alpha\beta}$ is given by

\be\label{t1} g_{\alpha\beta}=\eta_{CD}e^{C}_{\alpha} e^{D}_{\beta},\ee
\noindent
where $\alpha$ and $\beta$ are coordinate indices on the manifold and also run over $0, 1, 2, 3$, and $e^{C}_{\alpha}$ forms the tangent vector on the tangent space over which the metric $\eta_{CD}$ is defined. The non-null torsion $T^{\rho}_{\alpha\beta}$ and contorsion $K^{\alpha\beta}_{\rho}$ of Weitzenbock connection in Teleparallelism \cite{W} are defined by

\be\label{t2} T^{\rho}_{\alpha\beta} \equiv e^{\rho}_{C}[\partial_{\alpha}e^{C}_{\beta}-\partial_{\beta}e^{C}_{\alpha}] ,\ee
\be\label{t3} K^{\alpha\beta}_{ \rho} \equiv  -\frac{1}{2}[{T^{\alpha\beta}}_{\rho}-{T^{\beta\alpha}}_ {\rho}-{T_{\rho}}^{\alpha\beta}],\ee
\noindent
respectively. Moreover, instead of the Ricci scalar $R$ for the Lagrangian density in `General theory of Relativity', the Teleparallel Lagrangian density is presented by the torsion scalar $T$ as follows
\be\label{t4} T \equiv {S_{\rho}}^{\alpha\beta} {T^{\rho}}_{\alpha\beta},\ee
where,
\be\label{t5} {S_{\rho}}^{\alpha\beta} \equiv  \frac{1}{2}[{K^{\alpha\beta}}_{\rho}+{\delta}^{\alpha}_{\rho}{T^{\theta\beta}}_ {\theta}-{\delta}^{\beta}_{\rho}{T^{\theta\alpha}}_{\theta}].\ee

\noindent
The modified Teleparallel action of $F(T)$ gravity is given by

\be\label{t6}\mathbb{ A} = \int  d^4 x  \mid e \mid  F(T)+ S_m ,\ee
where $|e|$ = det $e^{C}_{\alpha}=\sqrt {-g}$, $S_m$ is the matter action, and the units has been chosen so that $c = 16 \pi G = 1$. Now, restricting ourselves to the spatially flat Robertson-Walker (R-W) space-time \footnote{Currently, there is a trend to refer the unique homogeneous and isotropic metric $ds^2 = -dt^2 + a^2\left[{dr^2\over 1-kr^2} + r^2 d\theta^2 + r^2 \sin\theta^2 d\phi^2\right]$ as FLRW metric. This is not correct, since model and metric are different. It is called Robertson-Walker line element, because Robertson (1935,1936) and Walker (1936) gave independent proofs that this particular line element describes the most general homogeneous and isotropic space-time geometry (this is found in any standard text book of GTR, e.g. `Gravitation' by Misner, Thorne and Wheeler, foot note pp 722). Independent derivations of evolving (decelerating) homogeneous and isotropic cosmological models were given by A. Friedmann (1922) and G. Lemaitre (1927) (Ibid, page 758). Therefore, the standard model of cosmology should be referred to as FLRW model, while we are considering deviation from the standard model with the unique isotropic and homogeneous R-W metric.}, described by,

\be\label{t6.1} {ds}^2 = - {dt}^2 + {a^2(t)} {dX}^2,\ee
where $a(t)$ is the scale factor, one finds $T = -6 {\dot a^2\over a^2}$. It is therefore customary to treat $T + 6 {\dot a^2\over a^2}=0$ as a constraint and introduce it in the action (\ref{t6})  through a Lagrange multiplier ($\lambda$) as,

\bes\label{t7}\mathbb{A}  = 2{\pi}^2\int\Big[F(T) - \lambda\Big\{T + 6\Big({\dot a^2\over a^2} \Big)\Big\} - {1\over 2}\phi_{,\mu}\phi^{,\mu} - V(\phi)\Big]a^3 dt,\end{split}\ee
where, $2\pi^2$ is an outcome of integration over the $3$-space. As already mentioned, $F(T)$ gravity has been introduced to drive late-stage of cosmic acceleration without the need for dark energy. A scalar field ($\phi$) has therefore been introduced in the above action \eqref{t7} to drive inflation at the very early stage of cosmic evolution \cite{3M}. Now varying the action with respect to $T$ one gets $\lambda = F'(T)$, where $F'(T)$ is the derivative of $F(T)$ with respect to $T$. Substituting it in the action one obtains,
\bes\label{t8} \mathbb{ A} = 2{\pi}^2\int\Big[F(T) - F'(T)\Big\{T + 6\Big({\dot a^2\over a^2} \Big)\Big\} + {1\over 2}\dot\phi^2 + V(\phi)\Big]a^3 dt .\end{split}\ee
Thus, the action finally  may be expressed in R-W metric (\ref{t6.1}) as

\be\label{t9} \mathbb{A}(a,\dot a,T,\dot T) = \int\left[-6a \dot a^2 F'+ a^3(F-F'T) + a^3 \left({1\over 2}\dot\phi^2 - V(\phi)\right)\right]dt,\ee
where we have absorbed the constant $2\pi^2$ in the action. At this stage let us mention that, for a general curvature parameter $k = 0 \pm 1$, associated with flat, closed and open universe, Ferraro and Fiorini \cite{4} computed appropriate vierbeins, and correspondingly obtained $T = 6\left[-H^2+{k\over a^2}\right]$. Therefore the field equations are,

\be \label{FE}\begin{split} & 12H^2 F'(T)+F(T)=16 \pi G\rho;\\&
4\left({k\over a^2} + \dot H\right)(12H^2 F'' (T)+F' (T))-F(T)-4F' (T)(2\dot H + 3H^2)=16\pi G p,\end{split}\ee
where, $\rho$ and $p$ are the energy density and thermodynamic pressure of a barotropic fluid respectively. However, since our present analysis is not affected by the presence of nonzero $k$, therefore we consider flat space $(k = 0)$, to avoid unnecessary complications.\\

Since, we focus on the evolution of the very early universe, we need to compute the Hamiltonian for canonical quantization. Clearly, unlike $F(R)$ gravity, the action \eqref{t9} is not canonical, since the Hessian determinant of the above action, being devoid of $\dot T$ term, vanishes and hence the action is singular. It is therefore required to perform Dirac's constraint analysis to find the canonical Hamiltonian. General Hamiltonian formulation of $f(T)$ gravity has already been performed by several authors \cite{HC1,HC2,HC3,HC4}. It has been found that due to the violation of Lorentz invariance, three extra degrees of freedom appear in $4$-dimensions \cite{HC1,HC2,HC3}. However, the issue is debatable \cite{HC4}. Nonetheless, around flat RW space-time, which is our present concern, $f(T)$ does not seem to propagate any additional degrees of freedom \cite{HC4}. \\

In the following section, we perform Dirac's constraint analysis, to compute canonical Hamiltonian for the $F(T)$ theory in the background of spatially flat R-W metric \eqref{t6.1} under consideration. The Hamiltonian so obtained is found to be impossible to handle, due to the presence of quadratic momentum in the denominator. Next, in section 3, we consider a particular form of $F(T)$, express the action as $\mathbb{A}(a,\dot a)$. The action thus becomes non-singular. It is important to mention that, the field equations of $F(T)$ gravity are second order, as in the case of Lanczos-Lovelock gravity. This is treated as the main advantage over $F(R)$ gravity, in which field equations are of fourth order or even higher. We show that, it is essentially a disadvantage, since it leads to the pathology of branched Hamiltonian, which has no unique resolution till date. The pathology is finally bypassed in section 4, taking into account $F(R,T)$ theory of gravity, by adding a curvature squared term ($R^2$) in the action. Semiclassical wavefunction has also been found in the subsection 4.1. Finally in section 5, we summarize the issue and our findings.

\section{Constraint analysis and the Hamiltonian:}

As mentioned, the Hessian determinant for the action \eqref{t9} vanishes, since $\dot T$ is not invertible. Note that the situation is different for $F(R)$ theory, since under integration by parts, $\dot R$ appears in the action, which is invertible. So in this section we sketchily analyse the constraint following Dirac's algorithm. The generic momenta with respect to variables $a$,$\phi$ and $T$ are:

\be\label{mom1} p_a = -12a\dot a F',~~~ p_\phi = a^3\dot \phi,~~~ p_T = 0.\ee
Clearly the constraint,

\be\label{C1} \xi = p_T \approx 0,\ee
vanishes weakly, since ${\partial \xi\over \partial p_T} \ne 0$. So, the constrained Hamiltonian reads as,

\be \label{Hc}H_c(a, T, \phi, p_{a}, p_{T}, p_\phi) = -\frac{p_{a}^2}{24a{F'}} + {{p_\phi}^2\over 2 a^3} - a^3(F-F'T)+ V(\phi)a^3 .\ee
The primary Hamiltonian may therefore be expressed in the following form,

\be\label{Hp} H_{p1} = H_c + \lambda p_T = -\frac{p_{a}^2}{24a{F'}} + {{p_\phi}^2\over 2 a^3}- a^3(F-F'T) + V(\phi)a^3 + \lambda p_T.   \ee
where, $\lambda$ is a Lagrange multiplier, and the Poisson brackets, $\{a, p_a\} = \{\lambda, p_T\} = 1$ hold. Since, $\{\xi, H_c\}$ does not vanish even weakly, so $\xi$ is a second class primary constraint. As the theory is devoid of first class primary constraint, so there must not exist any undetermined Lagrange multiplier of the theory. Now the constraint must be preserved in time, i.e.,

\be \dot \xi = \{\xi, H_{p1}\} = -F''\left({p_a^2\over 24a F'^2}  + a^3 T\right)\approx 0.\ee
Thus ($ \because F''\ne 0$),
\be\label{chi} \chi = {p_a^2\over 24a F'^2} + T a^3 \approx 0,\ee
is again a second class constraint, since

\be \label{pc}\{\xi, \chi\} = {p_a^2 F''\over 12 a F'^3} - a^3 \ne 0,\ee
although, it is not a new one. In fact, plugging in $p_a^2$ from  \eqref{mom1}, one can retrieve $T + 6{\dot a^2\over a^2} = 0$.
Next, since $\{\xi, \chi\} \ne 0$, so the constraint is second class as mentioned, and one has to make the consistency check by finding,

\be \begin{split}\label{cdot}\dot \chi = \{\chi, H\} &= \left(-{p_a^2\over 24 a^2 F'^2} + 3a^2 T\right)\left(-{p_a\over 12 a F'}\right)\\&-{p_a\over 12a F'^2} \left({p_a^2\over 24 a^2 F'} -{3\over 2}{p_{\phi}^2\over a^4}-3a^2(F - F'T) + 3V a^2\right) + \lambda\left[- {p_a^2 F''\over 12 a F'^3} + a^3\right],\end{split}\ee
which determines the lagrange multiplier as,

\be \lambda = {3 a^2 F'p_a\over p_a^2 F''- 12 a^4 F'^3}\left[F- 2F'T + {p_{\phi}^2\over 2 a^6} - V(\phi)\right].\ee
This results in the following form of the canonical Hamiltonian,

\be \label{H}H =  -\frac{p_{a}^2}{24a{F'}} - a^3(F-F'T) + {3 a^2 F'p_ap_T\over p_a^2 F''- 12 a^4 F'^3}\left[F- 2F'T + {p_{\phi}^2\over 2 a^6} - V(\phi)\right] .\ee
Clearly the Hamiltonian is extremely difficult to handle, if not impossible, particularly, since it contains momentum in the denominator. Note that in the absence of the scalar field, the Hamiltonian takes a simple form provided $2F'T = F$. But then, it simply means $F \propto \sqrt T = i\sqrt 6 {\dot a\over a}$, which is meaningless. Let us therefore consider a particular case.

\section{$F(T) = \beta T + \gamma T^2$:}

It is noteworthy that once a form of $F(T)$ is chosen, and the form of $T = -6{\dot a^2\over a^2}$ is substituted, the Lagrangian $L = L(a, \dot a. \phi, \dot\phi)$ becomes non-singular (the Hessian determinant is non-vanishing), and thus constraint analysis is no longer required. Regarding the above form of $F(T)$ we recall that the form  $F(T ) = (T^2 + 6\beta T - 3\beta^2))$ \cite{2d} and $F(T) = T + T^2 - c$ \cite{2e} are outcome of reconstruction program, which is the simplest generalization of $F(T)$ gravity. Further, the form $F(T) = \alpha T + \beta T^2$ has already been treated by several authors \cite{3M,2f,2g}, particularly to study early inflation and late stage of cosmic acceleration. In this section, we therefore proceed to construct the Hamiltonian in view of this form of $F(T)$. The point Lagrangian for the above form of $F(T)$ reduces to:

\be\label{PL1} L = -6\beta a \dot a^2 + 36\gamma{\dot a^4\over a} + a^3 \left({1\over 2}\dot \phi^2 - V(\phi)\right).\ee
The momentum canonically conjugate to the scale factor therefore is

\be p_a = -12\beta a\dot a + 144\gamma {\dot a^3 \over a}.\ee
Alas, the presence of cubic kinetic term in the momentum makes the theory intrinsically nonlinear, as in the case of Lanczos-Lovelock gravity theory. Hence the standard Hamiltonian formulation of such an action following the conventional Legendre transformation remains obscure. This pathology arises due to the fact that the Lagrangian is quartic in velocity, and therefore the expressions for velocity is multi-valued functions of momentum, resulting in the so-called multiply branched Hamiltonian (Energy) with cusps. This makes the classical solution unpredictable, as at any instant of time one can jump from one branch of the Hamiltonian (Energy) to the other, because the equation of motion allows for such jumps. Despite serious attempts \cite{BH1, BH2, BH3, BH4, BH5, BH6}, there is no unique resolution to this issue of branched Hamiltonian. For example, in view of a toy model, it was shown that, in the path integral formalism one can associate a perfectly smooth quantum theory possessing a clear operator interpretation and a deterministic classical limit \cite{BH1}. Nevertheless, it cannot be extended in the case of a realistic model. Further, it puts up question on the standard classical variational principle and on the canonical quantization scheme. Some authors tried to handle the issue, tinkering with some fundamental aspects, e.g., loosing Heaviside function to obtain manifestly hermitian convolution \cite{BH2}, sacrificing the Darboux coordinate to parametrize the phase space \cite{BH3} and ignoring the usual Heisenberg commutation relations \cite{BH4}. On the contrary, to obtain a single-valued Hamiltonian, Legendre–Fenchel transformation method was applied by some authors \cite{BH5}, while a modified version of Dirac’s constrained analysis following generalized Legendre transformation was employed by some others \cite{BH6}. It is noteworthy that although both \cite{BH5} and \cite{BH6} considered the same toy model, they happened to find two completely different Hamiltonians, which are not related under any sort of transformation \cite{BH7}. Thus, the pathology of branched Hamiltonian remains unresolved over decades.

\section{$F(R,T)$ theory of gravity:}

As mentioned, $F(T)$ gravity theory suffers from the pathology of branched Hamiltonian, and despite attempts over several decades, there is no unique resolution to the issue. Earlier, we have handled such situation in the context of Lanczos-Lovelock gravity theory \cite{BH7, BH8} as well as in Pais-Uhlenbeck oscillator action \cite{BH9}. It was shown that the problem although cannot be alleviated, but may be bypassed by adding a higher-order term in the action \cite{BH7, BH8, BH9}. In line with such earlier attempts, we therefore modify the $F(T)$ gravity by $F(R,T)$ gravity and express the action as,

\be \label{Action} A = \int \left[\alpha R^2 + \beta T +\gamma T^2 \right]\sqrt{-g}~d^4 x + \Sigma_{R^2},\ee
where, the supplementary boundary term $\Sigma_{R^2} = 4\alpha\int R K\sqrt h d^3x$, $K$ being the trace of the extrinsic curvature tensor $K_{ij}$, and $h$ is the determinant of the three metric $h_{ij}$. Since rest of our analysis is independent of the presence of a scalar field, and particularly because $R^2$ can drive inflation in the very early universe, while $T^2$ drives late-stage of cosmic acceleration, we have omitted the scalar field. In the flat R-W metric (\ref{t6.1}) under consideration,

\be \label{R,T} R = 6\left({\ddot a\over a}+{\dot a^2\over a^2}\right);\hspace{0.2 in}   T = -6 {\dot a^2\over a^2}.\ee
Now, in the modified Horowitz' formalism \cite{MH1,MH2,MH3,MH4,MH5,MH6,MH7,MH8,MH9,MH10,MH11}, the action in the first place, is expressed in terms of the basic variable $h_{ij} = a^2 \delta_{ij}= z\delta_{ij}$, so that $R$ and $T$ \eqref{R,T} may be expressed in terms of $z$ as,

\be \label{R,T,z} R = 3{\ddot z\over z};\hspace{0.2 in}   T = -{3\over 2}{\dot z^2\over z^2};\hspace{0.2 in} \sqrt{g} = z^{3\over 2},\ee
while the action reads as,

\be \label{Action1} A = \int\left[9\alpha{\ddot z^2\over \sqrt z} - {3\over 2}\beta {\dot z^2\over \sqrt z} + {9\over 4}\gamma {\dot z^4\over z^{5\over 2}}\right]dt + \Sigma_{R^2}.\ee
We now introduce the auxiliary variable

\be \label{Aux} q = {\partial A\over \partial \ddot z} = 18\alpha {\ddot z\over \sqrt z},\ee
and judiciously express the above action as,

\be A  = \int\left[q\ddot z -{\sqrt z q^2\over 36\alpha} - {3\over 2}\beta {\dot z^2\over \sqrt z} + {9\over 4}\gamma {\dot z^4\over z^{5\over 2}}\right]dt + \Sigma_{R^2}.\ee
Upon integration by parts, the total derivative term gets cancelled with the supplementary boundary term $\Sigma_{R^2}$, and we are left with

\be A = \int\left[-\dot q\dot z -{\sqrt z q^2\over 36\alpha} - {3\over 2}\beta {\dot z^2\over \sqrt z} + {9\over 4}\gamma {\dot z^4\over z^{5\over 2}}\right]dt.\ee
One can easily check that the definition of the auxiliary variable is recovered, from the $q$ variation equation. The canonical momenta are

\be\label{mom} p_q = -\dot z; \;\;\;\;\;p_z = -\dot q -3\beta {\dot z\over \sqrt z} + 9\gamma{\dot z^3\over z^{5\over 2}},\ee
and the Hamiltonian is expressed as

\be H = -p_q p_z + {3\over 2}\beta {p_q^2\over \sqrt z} - {9\over 4}\gamma{p_q^4\over z^{5\over 2}} + {\sqrt z q^2\over 36\alpha}.\ee
Now to translate Hamiltonian in terms of basic variables $\{h_{ij},K_{ij}\}$, let us make the transformation, viz. $q \rightarrow p_x$ and $p_q \rightarrow -x$, where, $x = \dot z$ (Bear in mind that $K_{ij} = -{1\over 2}\dot h_{ij} =  -{\dot z\over 2}\delta_{ij} = - {x\over 2}\delta_{ij}$). Note that the transformation is canonical since the Poisson bracket, $\{x,p_x\} = 1$ and else vanish. Thus the Hamiltonian reads as,

\be\label{PH1} H = x p_z + {\sqrt z\over 36\alpha}p_x^2 + {3\over 2}\beta {x^2\over \sqrt z} - {9\over 4}\gamma {x^4\over z^{5\over 2}} =0,\ee
which is constrained to vanish, due to diffeomorphic invariance.  This equation \eqref{PH1} is referred to as the classical Hamilton constraint equation. Canonical quantization of the above Hamiltonian leads to,

\be \label{Quant1}{i\hbar\over \sqrt z} {\partial \Psi\over \partial z} =-{\hbar^2\over 36 \alpha x}\bigg(\frac{\partial^2}{\partial x^2}+\frac{n}{x}\frac{\partial}{\partial x}\bigg)\Psi + {3\over 4}x\left[{2\beta \over z} - {3\gamma x^2\over z^3}\right]\Psi,\ee
where, the index $n$ removes some but not all the operator ordering ambiguities. The above modified Wheeler-DeWitt equation \eqref{Quant1}, under a further change of variable, $(z^{3\over 2} = \sigma)$, takes the following form,

\be\label{Quant2} i\hbar {\partial \Psi\over \partial \sigma} = -{\hbar^2\over 54 \alpha x}\bigg(\frac{\partial^2}{\partial x^2} +\frac{n}{x}\frac{\partial}{\partial x}\bigg)\Psi +  x\left[{\beta \over \sigma^{2\over 3}} - {3\gamma x^2\over 2\sigma^2} \right]\Psi = \hat H \Psi,\ee
where the proper volume $(\sigma = z^{3\over 2} = a^3)$ acts as the internal time parameter. The above Hamiltonian operator $\hat H$ is hermitian under the choice $n = -1$ \cite{MH6,MH7,MH8,MH9,MH10}, and hence the continuity equation,

\be {\partial \rho\over\partial \sigma} + \nabla.\mathbf{J} = 0,\ee
holds, where, $\rho = \Psi^*\Psi$, and $\mathbf{J} =(J_x,0,0)$ are the probability density and current density respectively, with $J_x = {i\hbar \over 54\alpha x}(\Psi\Psi_{,x}^* - \Psi^*\Psi_{,x})$. As a result, the standard quantum mechanical probabilistic interpretation holds. The effective potential $V_e = {1\over 2\sigma^2}\left(2\beta \sigma^{4\over 3} x - 3\gamma x^3\right)$ may be extremized with respect to $x$, to obtain

\be \label{deSitter} a = a_0 e^{\sqrt{\beta\over 18\gamma}t}, \ee
ensuring de-Sitter expansion.

\subsection{Semiclassical wavefunction:}

It is extremely difficult to find a solution to the modified Wheeler-DeWitt (W-D) equation \eqref{Quant1} or \eqref{Quant2}, due to tight coupling between the variables $x$ and $z~\mathrm{or}~(\sigma)$. However, the extremum of the effective potential obtained in de-Sitter form, motivates to study the behaviour of semiclassical wavefunction following an appropriate semiclassical approximation. The reason is, the semiclassical wavefunction clearly depicts the possibility  of transition from quantum to the classical domain with exponential de-Sitter expansion. It is noteworthy that, if the integrand in the exponent of the semiclassical wavefunction turns out to be imaginary, then the approximate wave function is oscillatory, and falls within the classical allowed region, otherwise, it falls within the classically forbidden domain. Particularly, the Hartle criterion \cite{HC} for the selection of classical trajectories states that: `if the approximate wavefunction obtained following some appropriate semiclassical approximation is strongly peaked around a classical solution, then there exists correlations among the geometrical and matter degrees of freedom, and the emergence of classical trajectories (i.e. the observable universe) is expected, on the contrary, if it is not peaked, correlations are lost’. Thus, Hartle criterion clearly plays the role of a selection rule to explore classical trajectories. In this subsection, we therefore study the behaviour of the modified W-D equation \eqref{Quant1}, following the standard WKB approximation, assuming an wavefunction in the form,

\be\label{s1} \Psi(x,z) = \Psi_0(x,z) e^{{i\over \hbar}S(x,z)},~~\mathrm{with}~~S(x,z) = S_0 + \hbar S_1 + \hbar^2 S_2+\cdots,\ee
where $S(x,z)$ is expanded as usual, in the power series of $\hbar$, and the prefactor $\Psi_0$ is a slowly varying function of $x$ and $z$. One can therefore compute,

\begin{center}
\be \label{s2} \begin{split} & \Psi_{,x}=\Psi_{0,x}e^{{i\over \hbar}S}+{i\over \hbar}\bigg[S_{0,x}+\hbar S_{1,x}+\hbar^2 S_{2,x}+\mathcal{O}(\hbar)  \bigg]\Psi_0 e^{{i\over \hbar}S};\\& \Psi_{,xx}=2{i\over \hbar}\bigg[S_{0,x}+\hbar S_{1,x}+\hbar^2 S_{2,x}+\mathcal{O}(\hbar)  \bigg]\Psi_{0,x} e^{{i\over \hbar}S}+{i\over \hbar}\bigg[S_{0,xx}+\hbar S_{1,xx}+\hbar^2 S_{2,xx}+\mathcal{O}(\hbar)  \bigg]\Psi_0 e^{{i\over \hbar}S}\\&+\Psi_{0,xx}e^{{i\over \hbar}S}- {1\over \hbar^2}\bigg[S_{0,x}^2+\hbar^2 S_{1,x}^2+\hbar^4 S_{2,x}^4+2\hbar S_{0,x} S_{1,x}+2\hbar^2 S_{0,x} S_{2,x}+2\hbar^3 S_{1,x}S_{2,x}+\mathcal{O}(\hbar)\bigg]\Psi_0 e^{{i\over \hbar}S} ;\\& \Psi_{,z}=\Psi_{0,z}e^{{i\over \hbar}S}+{i\over \hbar}\bigg[S_{0,z}+\hbar S_{1,z}+\hbar^2 S_{2,z}+\mathcal{O}(\hbar)  \bigg]\Psi_0 e^{{i\over \hbar}S}, \end{split}\ee
\end{center}
where,‘comma’ everywhere in the suffix represents derivative, and $\mathcal{O}(\hbar)$ stands for higher order terms. Now inserting the expressions (\ref{s1}) and (\ref{s2}) in the modified W-D equation (\ref{Quant1}) and equating the coefficients of different powers of $\hbar$ to zero, we obtain the following set of equations (upto second order),

\be\begin{split}
&\frac{\sqrt z}{36\alpha}S_{0,x}^2 + x S_{0,z} + \mathcal{V}(x,z) = 0, \label{s3}\end{split}\ee
\be\begin{split}
&-\frac{\sqrt z}{36\alpha}\bigg[ \bigg(iS_{0,xx}-2S_{0,x}S_{1,x}+{in\over x}S_{0,x}\bigg)\Psi_0+2iS_{0,x} \Psi_{0,x}\bigg]
+x S_{1,z}\Psi_0-ix\Psi_{0,z} = 0, \label{s4}\end{split}\ee
\be\begin{split}
&-\frac{\sqrt z}{36\alpha}\bigg[\bigg(iS_{1,xx}-S_{1,x}^2-2S_{0,x}S_{2,x}+{in\over x}S_{1,x}\bigg)\Psi_0+\Psi_{0,xx}+2iS_{1,x}\Psi_{0,x}+{n\over x}\Psi_{0,x}\bigg] + xS_{2,z}\Psi_{0}=0,
\label{s5}\end{split}\ee
where, $\mathcal{V}(x,z) = \left({3\beta x^2\over 2\sqrt z} - {9\gamma x^4 \over 4z^{5\over 2}}\right)$. First let us note that, identifying $S_{0,x}$ with $p_x$ and $S_{0,z}$ with $p_z$, the classical Hamiltonian constraint equation $H=0$, presented in equation \eqref{PH1}, is recovered straight away, from equation \eqref{s3}. Therefore, \eqref{s3} is identified as the Hamilton-Jacobi equation. We are now required to solve these above coupled set of differential equations \eqref{s3}-\eqref{s5} successively, to find $S_0$, $S_1$ and $S_2$ and so on, which is an extremely difficult task. However, we have already noticed that the extremum of potential \eqref{deSitter} is de-Sitter type. So, let us try to find the behaviour of the semiclassical wavefunction about classical de-Sitter solution, which under the replacements $z_0 = a_0^2$, reads as,

\be\label{s6} z = z_0 e^{\Lambda t},~\mathrm{so ~that}~~\dot z = \Lambda z = x,~\ddot z = \Lambda^2 z,\hspace{0.1 in}  where, \hspace{0.1 in}\Lambda = \sqrt{2\beta \over 9\gamma}.\ee
In view of the above relation between $x$ and $z$, one can express $S_{0, x} = {1\over \Lambda} S_{0,z}$. Hence the differential equation \eqref{s3} may now be solved to obtain $S_0$ as

\be\label{s7} S_0 = -12\alpha\Lambda^3\left[1 \mp \sqrt{1- {\beta\over 6\alpha \Lambda} + {\gamma\Lambda\over 4\alpha}}\right]z^{3\over 2} = -12\alpha\Lambda^3 \left[1\mp\sqrt{1 - {\gamma\over 2\alpha}}\right]z^{3\over 2}.\ee
The last term is realized upon substitution of $\Lambda = \sqrt{2\beta \over 9\gamma}$. The so obtained Hamilton-Jacobi function $S_0$ is ensured to be real under the condition $2\alpha > \gamma$, which essentially implies ${1\over 8\pi G} > \gamma$. Hence, the semiclassical wavefunction up to zeroth order approximation is found as,

\be\label{s8} \Psi = \Psi_0 e^{{i\over \hbar}\left[-12\alpha \Lambda^3\left(1\mp\sqrt{1-{\gamma\over 2\alpha}}\right)z^{3\over 2}\right]},\ee
which exhibits oscillatory behaviour.\\

\noindent
\textbf{First order approximation:}\\
Now, let us take up the first order approximation. Equation \eqref{s4} may be expressed as,

\be\label{s9} iS_{0,xx} - 2 S_{0,x}S_{1,x} + i{n\over x} S_{0,x} - {36\alpha x\over \sqrt z}S_{1,z} + 2i\left[{ S_{0,x}{\Psi_{0,x}\over \Psi_0}+{18\alpha x\over \sqrt z}}{\Psi_{0,z}\over \Psi_0}\right].\ee
Since $\Psi_0$ is a slowly varying function, so ${\Psi_{0,z}\over \Psi_0}$ and ${\Psi_{0,x}\over \Psi_0}$ terms, appearing in the above equation \eqref{s9} may be neglected for convenience. Now, under the identification $S_{0,x}$ with $p_x$ already made, and replacing $S_{1,z}$ by $S_{1,x}{dx\over dz} = \Lambda S_{1,x}$, the above equation \eqref{s9} may be expressed in terms of momentum and its derivative, as

\be \label{s10} i\left({p_{x,x}\over p_x} + {n\over x}\right) = 2S_{1,x}\left[1 + {18\Lambda\alpha x\over 18\Lambda^2\alpha z}\right] = 4S_{1,z},\ee
where, we have substituted $x = \Lambda z$ in view of \eqref{s6}, and expressed momenta $q = p_x = 18\alpha{\ddot z\over \sqrt z} = 18\Lambda^2\alpha \sqrt z$, in view of the definition of the auxiliary variable \eqref{Aux} respectively. Integration yields

\be\label{s11} S_1 = {i\over 4} \ln{(18\alpha\Lambda^{n+2}z^{{2n+1\over 2}})},\ee
where again we have substituted $p_x = 18\Lambda^2\alpha \sqrt z$. Therefore, upto first order approximation, one obtains,

\be\label{s12} \Psi(x,z) = \Psi_0  \left[18 \alpha \Lambda^{n + 2}z^{{2n+1\over 2}}\right]^{-{1\over 4}}e^{{i\over \hbar}\left[-12\alpha \Lambda^3\left(1\mp\sqrt{1-{\gamma\over 2\alpha}}\right)z^{3\over 2}\right]}.\ee
One can observe that first order approximation only affects the pre-factor of the semiclassical wavefunction, keeping the exponent unaltered. Likewise, higher-order approximations may be performed and the same oscillatory behaviour of the semi-classical wavefunction is administered. Since, the wave function is oscillatory about the classical de-Sitter solution, it implies that the wavefunction is strongly peaked around the classical de-Sitter (inflationary) solution. Therefore, according to Hartle prescription \cite{HC}, transition from quantum domain to the classical trajectory is confirmed. The universe thereafter enters the inflationary regime.

\section{Conclusion:}

Dark energy issue has puzzled cosmologists for over two decades, since despite tremendous effort, there is no evidence for the existence of a scalar field, as yet. This motivated scientists to find alternatives to the issue, following modifications of the geometric part of Einstein's `General Theory of Relativity'. $F(T)$ teleparallel gravity theory has been advocated for the same purpose, which was originally proposed by Einstein himself, in an attempt to unify gravity with electromagnetism. However, it is required to explore all such alternative proposals from different perspective, for validation. It has been shown by several authors, as mentioned in the introduction, that $F(T)$ gravity theory successfully can act as an alternative to the dark energy issue. It can also probe inflation successfully, in the presence of a scalar field, ensuring excellent fit with the currently released inflationary data \cite{AS}. It is therefore left to explore its role in the very early universe, particularly in the quantum domain. In the absence of a complete quantum theory of gravity, one studies quantum cosmology to get certain insights, as to what might have happened beyond Planck's scale, when every interaction including gravity is quantized. In the present manuscript, we therefore attempt to survey the role of $F(T)$ teleparallel gravity theory in this context, which requires canonical formulation of the theory, as a precursor.\\

The action for $F(T)$ teleparallel gravity is singular, and so it is required to analyse the constraint to cast the canonical Hamiltonian. This we have performed in the background of isotropic and homogeneous Robertson-Walker space-time. The canonical Hamiltonian so found, contains momenta in the denominator, and therefore it is impossible to handle. We therefore switch over to a particular form of $F(T) = \alpha T + \beta T^2$, that has been obtained earlier by several authors, under reconstruction programme. In the process, we reduce the configuration space variables $({a, T, \dot a, \dot T})$ to $({a, \dot a})$. However, such an action exhibits the very unpleasant fact that the theory is plagued with the pathology of branched Hamiltonian. This is an age old problem \cite{BH1}, and appears when kinetic term is present in the action with cubic or higher degree. For example, In the context of Lanczos-Lovelock gravity theory, the presence of cubic kinetic terms and quadratic constraints make the theory intrinsically nonlinear \cite{FD}. Even its linearized version is cubic rather than quadratic. Such a pronounced exotic behaviour of the action does not allow Hamiltonian formulation of Lanczos–Lovelock gravity following conventional Legendre transformation. Clearly, such notorious situation arises automatically if the Lagrangian is quartic in velocities. Since, in that case, the expression for velocities are multivalued functions of momentum, resulting in the so called multiply branched Hamiltonian with cusps. As a result, at any time one can jump from one branch of the Hamiltonian to the other, making classical solutions unpredictable. Further, the momentum does not provide a complete set of commuting observable, which results in non-unitary time evolution of quantum states. There is no unique resolution to the problem, despite attempts over decades. In this sense, the so-called main advantage of $F(T)$ gravity over $F(R)$ gravity theory, that the field equations are second order instead of fourth or higher, turns out to be a severe disadvantage, in particular.\\

However, we have earlier encountered this problem in the context of Lanczos-Lovelock gravity and Pais-Uhlenbeck oscillator action, and demonstrated a technique to bypass the issue, taking into account a higher-order term in the action. For gravity theory, a scalar curvature squared term in the action, suffices to bypass the pathology. It is important to mention that, in fact, all the quantum theories of gravity constructed so far, namely , different versions of string theories, supersymmetry and supergravity theories, contain curvature squared terms $(R^2,~R_{\mu\nu}R^{\mu\nu})$ in the effective action, under weak field approximation. It is noteworthy that in the background of isotropic and homogeneous RW metric, $R_{\mu\nu}R^{\mu\nu} - {1\over 3} R^2$ is a total derivative term, and so, it is enough to consider either. Therefore, here again the problem of branching has been resolved in the same manner, considering such an additional term $(R^2)$ in the action. The action $A = \int \left[\alpha R^2 + \beta T +\gamma T^2 \right]\sqrt{-g}~d^4 x$ is quite healthy, since $R^2$ term drives inflation in the early universe while $T^2$ drives late-stage of accelerated cosmic evolution, as demonstrated by several authors earlier. However, note that, at the late stage of cosmic evolution, if we seek solution in the form $a\sim t^n$, with $n > 1$ to assure accelerated expansion, then the contributions from $R^2 \sim 2n(2n-1)t^{(3n-4)}$, and that from $T^2 \sim 16 n^4 t^{(3n-4)}$, in view of the action \eqref{Action1}, appear with same power in `$t$'. Further, the condition $\alpha > 2\gamma$ is required to ensure real value of Hamilton-Jacobi function $S_0$ \eqref{s7}. Thus, both the terms contribute identically at the late stage of cosmic evolution. In fact, even if the standard exponential expansion $a \sim e^{\lambda t}$ or intermediate inflationary solution ($a\sim e^{At^f}, 0< f < 1, A > 0$) \cite{AS1, AS2, AS3} are sought, again both the terms are found to contribute identically. Therefore in no way one can ignore contribution from $R^2$ term at the late stage of cosmic evolution. \\

Summarily, in the present analysis with $F(R,T)$ model, the Hamiltonian so found is hermitian and the standard quantum mechanical probabilistic interpretation holds. Further, as a byproduct, we have found de-Sitter expansion upon extremization of the effective potential. Finally, we have performed a reasonably viable semiclassical approximation and notice that the wavefunction oscillates about the classical de-Sitter solution. This indicates possibility of transition to the classical inflationary regime. $F(R,T)$ gravity therefore appears to be more fundamental than $F(T)$ theory of gravity. By and by, we take the opportunity to mention that symmetric teleparallel gravity theory $F(Q)$ also suffers from the same problem of branching, which may be alleviated following the same technique.

\end{document}